\documentclass[twocolumn,showpacs,preprintnumbers,amsmath,amssymb]{revtex4}
\usepackage{graphicx}
\usepackage{dcolumn}
\usepackage{bm}
\begin{document}

\title{Energetics of holes trapped in DNA}
\author{Vadim Apalkov$^{\ast\dag}$ and Tapash Chakraborty$^{\ast\ddag}$}
\affiliation{$^\ast$Department of Physics and Astronomy,
University of Manitoba, Winnipeg, Canada R3T 2N2}
\affiliation{$^\dag$Department of Physics and Astronomy, Georgia 
State University, Atlanta, Georgia 30303, USA}
\date{\today}
\begin{abstract}
We report on our study of the electronic properties of guanine traps
in the DNA surrounded by adenines. We have shown that for a typical range 
of DNA parameters, formation of the bound state of two holes at the same 
guanine trap is possible for the GGG and GGGG traps if the hole-hole 
interaction is weak, which can be achieved for the DNA in solutions. The 
origin of the two-hole bound state is the competition between the Coulomb 
repulsion and the phonon mediated attraction between the holes. For the 
hole-phonon coupling constant $\approx 1$ two holes will be at the same 
trap if the on-site hole-hole repulsion energy is $\lesssim 0.9 $ eV.
\end{abstract}
\maketitle

The motion of a hole on DNA is usually studied by detecting its
trapping at a series of guanine sites incorporated in the DNA
\cite{book}. The hole can be injected into DNA through oxidation
process. Oxidation occurs under UV irradiation and
in the presence of certain oxidants. Since the guanine
has the lowest oxidation potential \cite{seidel96,steenken97},
the process takes place mainly at the guanine pair, which
results in the formation of a guanine radical cation.
A typical value for the oxidation potential difference $\Delta_{GA}$
for the G-pair and the A-pair is about 0.1-0.5 eV \cite{seidel96,steenken97},
which is also in the range of reported numerical estimates \cite{brunaud03}.
Finally, after the oxidation process takes place we get a guanine radical 
cation, or a hole injected into DNA at the G-pair. The next step then is the 
migration of the positive charge (hole) to the spots with lower energy 
\cite{saito98,lewis00} which are the sequence of many guanines, e.g. GG 
or GGG. The importance of this migration is based on the fact that
accumulation of holes at G-sequences makes these
spots vulnerable to mutations.

There are a few established pictures of hole migration through DNA
until it reaches the spot with the lowest energy: (i) Direct hole 
tunneling between G and GG (or GGG) spots. This transport mechanism 
works only when the distance between the G and GG pairs is small 
(a few base pairs). The transport rate then depends exponentially on the
transport distance \cite{geise00,geise01}.
(ii) Hole migration through subsequent tunneling:
This picture is applicable when the distance between
the G and GG spots is large (about 10-15 base pairs) and
when between the G spot, where the hole was originally injected, 
and the GG (or GGG) spots, there are a few G-pairs. In this case 
the hole migration is via subsequent tunneling between the
G-pairs \cite{geise00}. Since this transport process
is duffusive in nature, the transport rate has a power dependence on 
the distance between the G and GG spots.
(iii) The polaronic hopping transport: This picture works when between 
the G and GG (GGG) spots there are only A-pairs. When the distance 
between G and GG is small the transport mechanism is again via
tunneling (case (i)). However, for large distances, the hole at the 
G-spot tunnel into the nearest A base and then migrate via polaron 
hopping (diffusion) through the A$\ldots$A sequence to the GG 
(or a GGG) spot \cite{geise01}. The transport rate has a power 
dependence on the distance between G and GG.

Finally, the hole will be localized at the spots with the
lowest energy, i.e. GG or GGG or GGGG (whichever is present in the DNA).
A natural question that arises then is, if instead of a singe hole many
holes are injected into the DNA, what will be the final configuration
of the system? Which positions of the holes will have the lowest
energy? Can the GG, GGG, or GGGG spots trap two holes simultaneously?
In this case such spots will be highly reactive and vulnarable to mutation.
Formation of the bound state of two holes trapped by G-sites
is analogous to bipolaron formation in the homogeneous 1D
system \cite{mott94}. In this paper, we report on the properties of the 
G-traps, surrounded by adenines. The main question we are addressing 
here is that under what condition the G-traps can accumulate multiple 
holes.

The Hamiltonian of the hole system consists of three parts: (i) the
tight-binding Hamiltonian which includes the hole hopping between the
nearest base pairs and on-site energies of a hole, (ii) the hole-hole 
interaction Hamiltonian, and (iii) the Holstein's phonon Hamiltonian 
with diagonal hole-phonon interaction \cite{holstein}
\begin{equation}
{\cal H} = {\cal H}_{t} + {\cal H}_{i} + {\cal H}_{ph},
\label{H}
\end{equation}
where
\begin{equation}
{\cal H}_{t} = \sum_{i, \sigma} \epsilon_i a^{\dagger}_{i,\sigma}
a_{i,\sigma} - t \sum_{i,\sigma } \left[ a^{\dagger}_{i,\sigma}
a_{i+1,\sigma}  +h.c. \right]  ,
\label{Ht}
\end{equation}

\begin{equation}
{\cal H}_{i} = \sum_{i,j, \sigma} V_{i,j}n_{i,\sigma}
n_{j,-\sigma} +\sum_{i,j\neq i, \sigma}
 V_{i,j} n_{i,\sigma} n_{j,\sigma } ,
\label{Hi}
\end{equation}

\begin{equation}
{\cal H}_{ph} = \hbar \omega \sum_{i} b^{\dagger }_i b_{i}
+ \chi \sum _{i, \sigma} a^{\dagger}_{i,\sigma} a_{i, \sigma}
\left( b^{\dagger}_{i} + b_{i} \right),
\label{Hph}
\end{equation}
where $a_{i, \sigma}$ is the annihilation operator of hole with
spin $\sigma$ on site $i$, $\epsilon _i$ is the on-site
energy of hole, $b_i$ is the annihilation operator of
phonon on site $i$, $t$ is the hopping integral between the
nearest base pairs (sites), $\omega $ is the phonon frequency,
$\chi $ is hole-phonon coupling constant, and $n_{i,\sigma}=a^\dagger
_{i,\sigma}a_{i,\sigma}$. The Hamiltonian [Eq.~(\ref{H})] without
the ${\cal H}_{ph}$ was studied earlier for a homogeneous system
in Ref.~\cite{vadim}.

In the tight-binding Hamiltonian [Eq.~(\ref{Ht})], we assume that
the site $i$ can be either a adenine or a guanine. We then take
the energy of the hole on the A site as zero energy, i.e. $\epsilon_A= 0$,
and the energy of the hole on the G-site to be negative,
$\epsilon_G=-\Delta _{GA}<0$. In the interaction Hamiltonian ${\cal H}_i$,
we take into account only the Hartree interaction
between the holes. The first term in Eq.~(\ref{Hi}) describes the
repulsion between the two holes with different spin. The holes
can then occupy the same site. The second term in Eq.~(\ref{Hi})
corresponds to two holes with the same spin. To get the basic idea 
about the typical range of interaction parameters when many holes are
trapped by the G-sites, we introduce a single-parameter interaction
potential of the form
\begin{equation}
V_{i,j} = \frac{V_0}{\sqrt{(i-j)^2+1}},
\label{Vij}
\end{equation}
where $V_0$ is the on-site repulsion between the two holes.
The origin of the form of $V_{i,j}$ in Eq.~(\ref{Vij}) is the finite 
spreading of the hole on-site states. This spreading is  
about the distance between the nearest base pairs.
Although the actual dependence of interaction potential on
the separation between the holes is more complicated \cite{starikov03}
than Eq.~(\ref{Vij}), this difference is not important for our
analysis since only the on-site interaction plays the
main role in formation of the bound state of two holes \cite{magna97}. 
In the hole-phonon Hamiltonian ${\cal H}_{ph}$, we 
include only the optical phonons \cite{bishop02} with diagonal
hole-phonon interaction, and do not take into account acoustic 
phonons which results in non-diagonal hole-phonon interaction 
\cite{conwell00,conwell01}, i.e. modify the tunneling integral.
In Eqs.~(\ref{Ht})-(\ref{Hph}) we also assumed that the hopping
integral $t$, the phonon frequency $\omega $, and the hole-phonon
coupling constant $\chi $, do not depend on the type of the base pairs 
(A or G).

The form of the total Hamiltonian, Eqs.~(\ref{H})-(\ref{Hph}),
leads to four dimensionless parameters which
characterize the system: The nonadiabaticity parameter \cite{magna96}
$\gamma = \hbar \omega /t$, the canonical hole-phonon coupling constant
\cite{magna96} $\lambda = \chi^2/(2\hbar \omega t)$, dimensionless hole-hole
interaction strength $V_0 /t$, and the dimensionless difference between
on-site energies of G and A, $\delta_{GA} = \Delta_{GA}/t$.

\begin{figure}
\begin{center}\includegraphics[width=7.6cm]{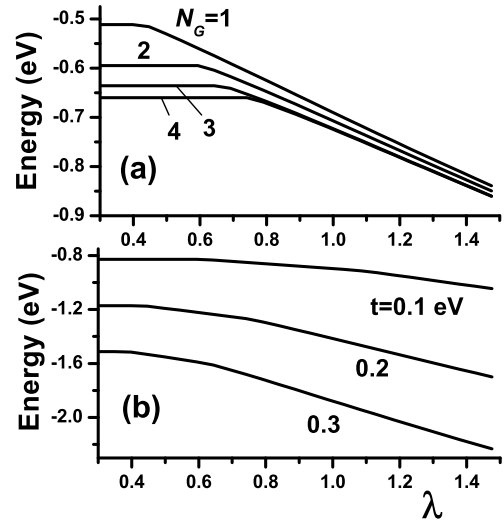}\end{center}
\vspace*{-1cm}
\caption{(a) The ground state energy of a single hole in a trap 
contaning $N_G$ guanines for $t=0.2$ eV and $\Delta_{GA}=0.3$ eV as a 
function of the hole-phonon coupling constant, $\lambda $. The numbers 
next to the lines are the number of guanines in the trap.
(b) The energy $E_{1,1}+E_{1,4}$  of two holes for $\Delta _{GA}=0.3$ eV 
as a function of $\lambda $. The numbers next to the lines are the values 
of tunneling integral, $t$. }
\label{figone}
\end{figure}

We numerically determine the eigenfunctions and eigenvectors of the
hole-phonon system by exactly diagonalizing the Hamiltonian
Eqs.~(\ref{H})-(\ref{Hph}) for a finite size system consisting of
six base pairs (sites). We also introduce limitations on the total
number of phonons \cite{marsiglio95},
$\sum_{i} n_{ph,i} \leq  N_{max}$,
where $n_{ph, i}$ is the number of phonons on site $i$.
To compare the energy spectrum of the systems with different number
of holes, we keep the maximum number of phonons per hole
the same for all systems. For the two-hole system the maximum number of
phonons is $N_{max}=16$ and for the one-hole system $N_{max}=8$.

\begin{figure}
\begin{center}\includegraphics[width=7.6cm]{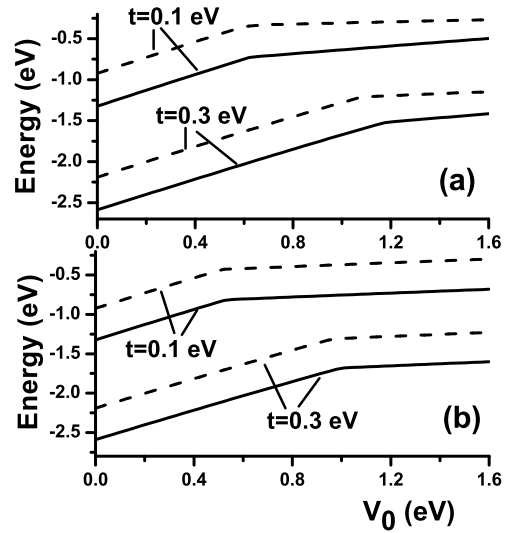}\end{center}
\vspace*{-1cm}
\caption{Ground state energy of two holes in the trap containing 
$N_G=3$ guanines (a) and $N_G=4$ guanines (b) as a function of the 
inter-hole ineraction strength, $V_0$, for $\Delta_{GA}=0.1$ eV 
(dashed line) and $\Delta_{GA}=0.3$ eV (solid line).
}
\label{figtwo}
\end{figure}

Our finite size system contains six sites which are originally
adenines. We then introduce the G-traps with different number $N_{G}$
of guanines, G, GG, GGG, and GGGG, in the middle of the system. 
For example the system with two guanines is AAGGAA. For different 
traps we calculate the energy of the ground state of the systems 
with one and two holes.  Denoting the corresponding energies as 
$E_{1,N_G}$ (for the one-hole system with $N_G$ guanines) and $E_{2, N_G}$ 
(for the two-hole system with $N_G$ guanines) we write the condition
that the trap with $N_{G}$ guanines will accomodate two holes as
\begin{equation}
E_{2,N_G} < E_{1,N_G}+E_{1,1}.
\label{trap2}
\end{equation}
The meaning of condition ($\ref{trap2}$) is as follows:
If two holes are injected into the DNA then the lowest energy
of the system corresponds to the case when the holes are
trapped by the same trap with $N_G$ guanines. The condition 
(\ref{trap2}) will determine the critical value of the hole-hole 
interaction strength, $V_{0}^{cr}$. That means for 
$V_0 < V_{0}^{cr}$ two holes will be trapped by the same trap with
$N_G$ guanines. For $V_0 > V_{0}^{cr}$ such a trapping is
energetically unfavorable and two holes will be at
different traps. 

As we shall see below the condition Eq.~(\ref{trap2}) is
not the condition for formation of bound state of two holes within
the trap containing $N_G$ guanines. The condition for formation
of bound state of two holes should give different values
for the critical hole-hole interaction strength, $V_{0}^{b}$.
The values $V_{0}^{cr}$ and $V_{0}^{b}$ would coincide only
for an infinite homogeneous system. In our case the system is finite,
which results in an inequality $V_{0}^{cr}< V_{0}^{b}$. The
finite system in our problem is actually the finite number of guanine
sites in the traps.

\begin{figure}
\begin{center}\includegraphics[width=7.6cm]{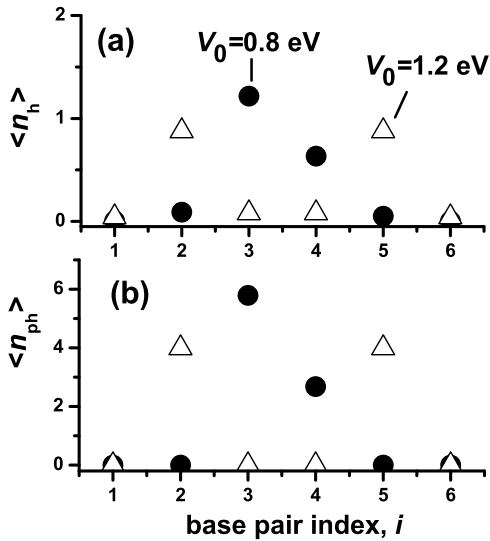}\end{center}
\vspace*{-1cm}
\caption{The average number of holes (a) and
the average number of phonons (b) for a two hole
system in a GGGG trap are shown as a function of the base index.
The tunneling integral is $t=0.3$ eV and the
hole-phonon coupling is $\lambda =1$. Dots and triangles
corresponds to inter-hole interaction strength
$V_0=0.8$ eV and 1.2 eV respectively.
}
\label{figthree}
\end{figure}

For our investigation of the system [Eqs.~(\ref{H})-(\ref{Hph})], we 
consider the following typical DNA parameters: $0.1$ eV $< t < 0.3$ eV, 
$0.1$ eV $< \Delta _{GA} < 0.5$ \cite{seidel96,hu02}, $0.05$ eV 
$< \hbar \omega < 0.1$ eV. For the dimensionless canonical hole-phonon 
coupling constant we have taken the value $\lambda=1$. For this coupling 
constant, the size of the polaron is about 2-3 base pairs. 
Our calculations show that the critical value $V_{0}^{cr}$ is very small
($V_0^{cr} \approx 0.1$ eV) when two holes have the same spin and they
can not occupy the same site. This small value of $V_0^{cr}$ also illustrate 
the fact that the phonon mediated attraction between the holes are largest 
when the holes occupy the same site. Therefore, in what follows we shall 
consider only the case of two holes with opposite spin.

In Fig.~1a the ground state energy of a single hole is plotted as a
function of the hole-phonon coupling constant, $\lambda $, for different
types of traps. For $\lambda \approx 1$ the difference
between the bound state of a hole in G and GG traps is about
0.03 eV, which is smaller than the value (0.05 eV) obtained in
Ref.~\onlinecite{basko01}. The size of the polaron in our calculations is
2-3 base pairs depending on the values of $t$ and $\omega$. As a next
step we calculate the ground state energy of two holes, $E_{1,N_G}+E_{1,1}$,
one of which is bound at a G trap and another one at a GGG or GGGG trap.
As an example, in Fig.~1b the ground state energy of two holes with the
second hole bound by GGGG is plotted as a function of
$\lambda $ for different values of the hopping integral, $t$.

\begin{figure}
\begin{center}\includegraphics[width=7.6cm]{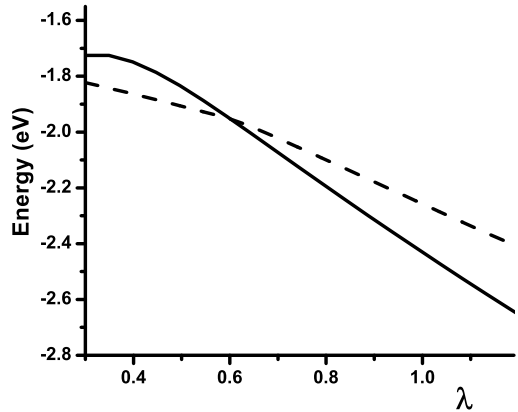}\end{center}
\vspace*{-1cm}
\caption{Energies $E_{1,1}+E_{1,4}$ and $E_{2,4}$
of a two-hole system are shown as a function of hole-phonon
coupling, $\lambda $, by dashed and solid lines, respectively.
Tunneling integral is $t=0.3$ eV and $\Delta_{GA}=0.3$ eV.
}
\label{figfour}
\end{figure}

In Fig.~2 the ground state energy $E_{2,N_G}$ of two holes bound in
a single trap is plotted for $N_G=3$ (Fig.~2a) and for $N_G=4$
(Fig.~2b) as a function of the hole-hole interaction strength for
different values of $\Delta_{GA}$ and $t$. Here we notice that
at some critical value $V_0^b$ of the hole-hole interaction strength,
there is a change of slope in the $E_{2,N_G}(V_0)$ dependence. This
critical value corresponds to the condition that the two holes are
bound in the G-traps, forming a bipolaron. The illustration of this
fact is given in Fig.~3. In Fig.~3a the average number of holes,
$\langle n_h\rangle=\langle n_{i,\sigma}\rangle+\langle 
n_{i,-\sigma}\rangle$, 
is shown as a function of the base pair index for a GGGG trap
and two different values of the hole-hole interaction strength, $V_0$.
It is clearly seen that for $V_0=0.8$ eV $< V_0^{b}$ the two holes are
almost at the same G sites, while at $V_0 = 1.2$ eV $> V_0^{b}$
the holes are away from each other. The corresponding distribution
of the average number of phonons $\langle n_{ph}\rangle$, is shown 
in Fig.~3b.

Another critical value of $V_0$ is introduced by equation
Eq.~(\ref{trap2}). The competition between $E_{1,1}+E_{1,N_G}$ and
$E_{2,N_G}$ is illustrated in Fig.~4. Comparing the energies
$E_{1,N_G}+E_{1,1}$ and $E_{2,N_G}$ for $\lambda =1$ and
different values of $t$, $\Delta _{GA}$, and $\omega $
one can determine $V_{0}^{cr}$. The result is
summarized in Table~1 for the GGGG trap. The corresponding results
for the GGG trap is about 0.1 eV smaller values for $V_0^{cr}$.
The dimensionless parameters, $\gamma $, $\delta_{GA}$, and
$V_0^{cr}/t$, are also given in Table~1. From this data
we can conclude that within the present range of parameters
the dependence of $V_0^{cr}$ on $\Delta _{GA}$ is weak, and
$V_0^{cr}/t$ depends mainly on $\gamma$. This dependence
can be approximated by a linear function as
\begin{equation}
V_0^{cr} \approx 3 \gamma t + 1.6 t \approx 3 \hbar \omega  + 1.6 t.
\label{linear}
\end{equation}

\begin{table}
\caption{Calculated values of $V_{0}^{cr}$ for various values of the
dimensionless DNA parameters}

\begin{tabular} {|c|c|c|c|c|c|c|} \hline

t (eV) & $\hbar\omega$ (eV) & $\Delta_{GA}$ (eV) & $V_0^{cr}$
(eV) & $\gamma$ & $\delta_{GA}$ & $V_0^{cr}$/t \\ \hline
0.1  &  0.1  & 0.1  &  0.42  & 1.00  &  1.00  &  4.2   \\
0.1  &  0.1  & 0.3  &  0.45  & 1.00  &  3.00  &  4.5   \\
0.1  &  0.1  & 0.5  &  0.46  & 1.00  &  5.00  &  4.6   \\
0.1  &  0.05 & 0.1  &  0.36  & 0.50  &  1.00  &  3.6   \\
0.2  &  0.1  & 0.1  &  0.42  & 0.50  &  0.50  &  3.25  \\
0.2  &  0.1  & 0.3  &  0.59  & 0.50  &  1.50  &  2.95  \\
0.2  &  0.1  & 0.5  &  0.62  & 0.50  &  2.50  &  3.1   \\
0.3  &  0.1  & 0.1  &  0.82  & 0.33  &  0.33  &  2.7   \\
0.3  &  0.1  & 0.3  &  0.78  & 0.33  &  1.00  &  2.6   \\
0.3  &  0.1  & 0.5  &  0.82  & 0.33  &  1.67  &  2.7   \\
0.3  &  0.05 & 0.3  &  0.60  & 0.17  &  1.00  &  1.97  \\
0.3  &  0.05 & 0.5  &  0.62  & 0.17  &  1.67  &  2.07  \\ \hline
\end{tabular}
\end{table}

We see from these data that depending on the parameters
of DNA, the critical hole-hole interaction strength, $V_0^{cr}$
can range from 0.4 eV to 0.8 eV. Numerical analysis
of the electron correlations in different types of DNA
\cite{starikov03} shows that the hole-hole interaction strength
is around 0.9 eV for A-DNA and  1.5 eV for B-DNA.
Additional suppression of the inter-hole interaction
by a factor of $\approx 0.7$ \cite{basko02}
can occur for DNA in solution, when hole-hole interacton is
screened by polar solvent molecules and mobile counterions.
Under this condition trapping of two holes by GGG and
GGGG traps would be possible.

In conclusion, we have shown that for a typical range of
DNA parameters, formation of bound state of two holes at the
same guanine trap is possible for GGG and GGGG traps if the
hole-hole interaction is sufficiently weak, which can be achieved
for DNA in solutions. Formation of the two-hole bound state results
in double positive charge of G-traps, which should modify their
chemical properties and increase the reactivity of these traps.
Experimental study of the two-hole states in G-traps should also
give additional insight on the hole-phonon interaction strength.
Formation of the bound state of two holes at the G-traps requires a
strong hole-phonon interaction, which should overcome the hole-hole
Coulomb repulsion. In our calculations the hole-phonon coupling constant 
was $\lambda = 1$ which is larger than the experimentally reported 
$\lambda \approx 0.2$ in Ref.~\cite{omerzu04}. Hence, experimental 
observation of the two-hole bound state should give an
additional estimate for the strength of hole-phonon interaction.

The work of T.C. has been supported by the Canada 
Research Chair Program and the Canadian Foundation for Innovation 
Grant.


\begin{thebibliography}{99}
\bibitem[\ddag]{byline} Electronic mail:
tapash@physics.umanitoba.ca

\bibitem{book}
{\it Long-Range Charge Transfer in DNA}, Ed. G.B. Schuster,
(Springer, New York 2004).

\bibitem{seidel96} C.A.M. Seidel, A. Schulz, and H.M. Sauer,
J. Phys. Chem. {\bf 100}, 5541 (1996).

\bibitem{steenken97}S. Steenken and S.V. Jovanovic, J. Am.
Chem. Soc. {\bf 119}, 617 (1997).

\bibitem{brunaud03} G. Brunaud, F. Castet, A. Fritsch,
and L. Ducasse, Phys. Chem. Chem. Phys. {\bf 5}, 2104
(2003).

\bibitem{saito98} I. Saito, T. Nakamura, K. Nakatani,
Y. Yoshioka, K. Yamaguchi, and H. Sugiyama, J. Am. Chem. Soc.
{\bf 120}, 12686 (1998).

\bibitem{lewis00} F.D. Lewis, X. Liu, J. Liu, R.T. Hayes,
and M.R. Wasielewski, J. Am. Chem. Soc. {\bf 122}, 12037 (2000).

\bibitem{geise00} B. Giese, Acc. Chem. Res. {\bf. 33}, 631 (2000).

\bibitem{geise01} B. Giese, J. Amaudrut, A.K. Kohler,
M. Spormann, and S. Wessely, Nature {\bf 412}, 318 (2001).

\bibitem{mott94} A.S. Alexandrov and N.F. Mott, Rep. Prog. Phys.
{\bf 57}, 1197 (1994).

\bibitem{holstein} T. Holstein, Ann. Phys. {\bf 8}, 325 (1959);
{\bf 8} 343 (1959).

\bibitem{vadim} V.M. Apalkov and T. Chakraborty, Phys. Rev. B {\bf 
71}, 033102 (2005).

\bibitem{starikov03} E.B. Starikov, Phil. Mag. Lett. {\bf 83},
699 (2003).

\bibitem{magna97} A. La Magna and R. Pucci, Phys. Rev.
B {\bf 55}, 14886 (1997).

\bibitem{bishop02} S. Komineas, G. Kalosakas, and A.R. Bishop,
Phys. Rev. E {\bf 65}, 061905 (2002).

\bibitem{conwell00} E.M. Conwell and S.V. Rakhmanova, Proc.
Natl. Acad. Sci. USA. {\bf 97}, 4556 (2000).

\bibitem{conwell01} S.V. Rakhmanova and E.M. Conwell, J. Phys.
Chem. B {\bf 105}, 2056 (2001).

\bibitem{magna96} A. La Magna and R. Pucci,
Phys. Rev. B {\bf 53}, 8449 (1996).

\bibitem{marsiglio95} F. Marsiglio, Physica C {\bf 244},
21 (1995). 

\bibitem{hu02} X. Hu, Q. Wang, P. He, and Y. Fang, Anal. Sci.
{\bf 18}, 645 (2002).

\bibitem{basko01} E.M. Conwell, D.M. Basko, J. Am. Chem. Soc.
{\bf 123}, 11441 (2001).

\bibitem{basko02} D.M. Basko and E.M. Conwell,
Phys. Rev. Lett. 88, 098102 (2002).

\bibitem{omerzu04} A. Omerzu, M. Licer, T. Mertelj, V.V. Kabanov,
and D. Mihailovic, Phys. Rev. Lett. {\bf 93}, 218101 (2004).

\end{thebibliography}
\end{document}